# Observation of multiple superconducting gaps in $Fe_{1+y}Te_{1-x}Se_x$ via a nano-scale approach to point-contact spectroscopy

Haibing Peng[*], Debtanu De, Zheng Wu, Carlos Diaz-Pinto

Department of Physics and the Texas Center for Superconductivity, University of Houston, 4800 Calhoun Rd, Houston, Texas 77204-5005

[*] Corresponding author: haibingpeng@uh.edu

ABSTRACT

We report a distinct experimental approach to point-contact Andreev reflection spectroscopy with *diagnostic capability* via a unique design of nano-scale normal metal/superconductor devices with excellent thermo-mechanical stability, and have employed this method to unveil the existence of two superconducting energy gaps in iron chalcogenide $Fe_{1+y}Te_{1-x}Se_x$ which is crucial for understanding its pairing mechanism. This work opens up new opportunities to study gap structures in superconductors and elemental excitations in solids.

Point-contact spectroscopy is an important tool not only for probing elementary excitations in solids, but also for studying superconducting gap structures via Andreev reflection (AR) at normal metal/superconductor (N-S) contacts. However, significant barriers remain to improve thermo-mechanical stability and provide diagnostic information on point contacts traditionally fabricated by mechanically pressing a metal tip against a bulk superconductor. Here we describe a distinct experimental approach to point-contact AR spectroscopy with *diagnostic capability* via a unique design of nano-scale N-S devices with excellent thermo-mechanical stability, and have employed this to unveil the existence of two energy gaps in $Fe_{1+y}Te_{1-x}Se_x$ which is crucial for understanding its pairing mechanism.

Fe-based superconductors have been attracting enormous interest [1-6] because of their high transition temperatures and intriguing physical mechanisms involving superconducting (SC) and magnetic orders.[7,8] The iron chalcogenide $Fe_{1+y}Te_{1-x}Se_x$, [6,9] with a simple crystal structure and multiple Fermi pockets, [10] provides a unique platform for investigating the pairing symmetry [11,12] in Fe-based superconductors. Related experiments have led to the observation of a sign-reversal of the SC gap function between the electron and hole pockets,[11] in line with a possible pairing mechanism with s± symmetry.[13] However, the number of SC gaps, another key issue for understanding the pairing, remains elusive. Experimental results based on traditional approaches [11,12,14-16] have been inconclusive and suffered from the lack of diagnostic details. Concrete evidence for multiple SC gaps has not been obtained in $Fe_{1+y}Te_{1-x}Se_x$ by scanning tunneling microscopy (STM) [11,12,16] or traditional point-contact spectroscopy,[15] while specific heat measurements have indicated the presence of two gaps.[14]

In N-S junctions with low potential barriers, a quantum transport phenomenon, known as Andreev reflection, occurs at the interface when the energy of incoming electrons is less than the SC gap energy. Point-contact AR spectroscopy **[17-22]** can provide fundamental information on the energy gap $\Delta$, since an enhanced electrical conductance occurs at bias voltage $V < \Delta / e$ across a N-S junction via the reflection of an incoming electron from the N side as a hole of opposite wave vector. The common method for fabrication of point contacts is the "needle-anvil" technique, where a metallic tip is mechanically pressed against a bulk superconductor. However, significant challenges remain to overcome the poor thermo-mechanical stability of thus-prepared N-S point contacts, the lack of control on the actual size of the point contacts due to the inevitable deformation of the metal tip, and the lack of important diagnostic information on the surface. In addition, the mechanical pressure from the tip can affect the SC properties at the junction. Diagnostic tools are thus critically needed to address the actual $T_c$ of the superconductor grain dominating a N-S junction. In particular, a degraded surface layer usually exists in Fe-based superconductors, and thus diagnostic information is indispensible for interpreting experimental results yet unfortunately not available in traditional AR methods. In this work, we describe a distinct experimental approach to point-contact spectroscopy by designing devices with suspended superconducting micro-scale crystals bridging multi-terminal normal metal electrodes (Fig. 1), and have employed this to unveil the existence of two energy gaps in $Fe_{1+y}Te_{1-x}Se_x$. AR spectroscopy is implemented in a three-terminal scheme (Fig. 1a) to address the target N-S junction, while the multi-terminal configuration allows diagnostic experiments for determining the actual $T_c$ and the conduction regime in the N-S junction. Furthermore, our approach offers excellent thermo-mechanical stability, provides a better control on the contact size and prevents the effect of mechanical pressure at the contact, thus enabling its application as a powerful and widely applicable spectroscopic tool.

In experiments, we start with single-crystal bulk materials of $FeTe_{0.5}Se_{0.5}$ (bulk $T_c$ ~ 14.2 K). After mechanically cleaving the bulk crystal into micro-meter scale crystals, we manipulate a microcrystal onto multi-terminal metal electrodes with a sharp probe tip attached to a micro-manipulator.[23] The optical image of a typical device is shown in Fig. 1b, where a SC microcrystal is suspended on top of six parallel electrodes spaced ~500 nm apart as designed via electron-beam lithography on $SiO_2$/Si substrates with a 200 nm thick $SiO_2$ layer. The critical temperature and the critical current of the SC microcrystal are characterized in a standard four-terminal configuration. To obtain AR spectra for a N-S junction, we have designed a circuit (Fig. 1a) where a small AC current superimposed to a DC bias current is driven between the $I_+$ and $I_-$ terminals while both the DC voltage drop and the AC voltage drop across the N-S junction are monitored between the $V_+$ and $V_-$ terminals. By this, the AR spectrum, i.e. the differential conductance *dI/dV* vs. the DC voltage bias *V*, is obtained for the target junction between the $I_+$ ($V_+$) electrode and the superconductor. We note that the $I_+$ and $V_+$ terminals are connected to the opposite ends of the same electrode so that the measured voltage drop is exactly across the target N-S junction by excluding the electrode resistance in series.

Fig. 2a presents the normalized *dI/dV* vs. *V* at *T* = 240 mK for the N-S junction between electrode 4 and the superconductor of the device shown in Fig. 1b. Our major findings include two characteristic features in the *dI/dV* spectrum: (1) two clear plateaus for |V| < 5 mV, and (2) a dip before the *dI/dV* returns to the normal state value at |V| ~ 15 mV. The Andreev reflection ratio at zero bias is ~ 1.7, indicating a nearly transparent N-S interface characterized by a weak barrier strength *Z* according to the BTK theory.[17] It is remarkable that the normal state conductance at high bias is the same for different temperatures (as shown by the raw data of *dI/dV* vs. *V* in Fig. 2b), demonstrating excellent thermo-mechanical stability important for obtaining correct normalized *dI/dV* for theoretical analysis. As temperature *T* is increased (Fig. 2b), the two plateaus at low bias shrink with the smaller plateau vanishing at *T* ~ 9.8 K and the larger one vanishing at *T* ~15 K. On the other hand, the position of the characteristic *dI/dV* dip shifts to lower bias as *T* increases, while the dip depth is reducing until it is not distinguishable above *T* ~ 12.7 K. Similarly, as magnetic field increases, the plateaus shrink although both of them are still observable at *B* = 14 T (Fig. 2c).

Next, we employ the unique diagnostic capability in our method to understand the characteristic features described above. First, we discuss the origin of the *dI/dV* dip preceding the normal state. Similar dips have been reported in traditional point-contact AR spectra for bulk superconductors, but they are not accounted for by the BTK theory and their physical origin is still under debate due to the lack of diagnostic details on the point contact.[20,24,25] Previously, the role of critical current $I_c$ in the junction was considered and the dips were suspected to be caused by point contacts not being in the ballistic limit.[25] Alternatively, a proximity effect [24] was proposed to explain the dips. Here, we show via diagnostic experiments that such *dI/dV* dips occur when the critical current $I_c$ of the corresponding section of the SC microcrystal is reached (Fig. 3a).

In our experiments [23], such a concurrent occurrence of the *dI/dV* dip and the critical current $I_c$ has been observed in many junctions. Based on this observation, the *dI/dV* dip can be best explained by a smooth change in I-V curves for a N-S junction from a higher linear conductance state at low bias *V* (due to Andreev reflection) to a lower linear conductance state at high bias (following Ohm's law after the normal state is reached), since a dip of *dI/dV* occurs naturally in the transition region where an inflection point must exist in order to connect the two linear I-V curves both extending through the origin (Fig. 3b).

In addition, due to the distinct multi-terminal device configuration, we are capable of obtaining AR spectra for the same N-S junction with different combinations of the $V_-$ and $I_-$ terminals. As seen in Fig. 3c, the dip position and its depth vary with different electrode configurations, since the electric field distribution and thus the critical current depend on the selection of the $I_+$ and $I_-$ terminals owing to an irregular point-contact geometry, consistent with our previous discussion on the role of $I_c$ in the origin of the *dI/dV* dip. However, the two center plateaus remain the same regardless of the selection of the $V_-$ and $I_-$ terminals, indicating that these two plateaus are related to intrinsic superconducting properties.

Further detailed analysis shows that the two plateaus are attributable to the existence of two SC gaps. The solid line in Fig. 2a represents a fitting of the experimental *dI/dV* data (normalized by the data at *T* = 15.4 K in Fig. 2b) by the generalized BTK theory considering the broadening effect.[26,27] The plateaus are well fit in a two-gap scenario by expressing the normalized conductance as $\sigma = w_1 \sigma_1 + (1 - w_1) \sigma_2$, where $\sigma_1$ and $\sigma_2$ are the normalized conductance for the two SC gaps, respectively, calculated according to the BTK theory,[26] and $w_1$ is the weight of contribution from the first gap. The fitting gives the energy of the SC gaps as $\Delta_1 = 0.63$ meV (with a broadening $\Gamma_1 = 0.10$ meV) and $\Delta_2 = 2.28$ meV (with a broadening $\Gamma_2 = 0.40$ meV), respectively, the weight of contribution for gap one is $w_1 = 0.34$ and the barrier strength is $Z = 0.14$. The obtained energy value of the second gap $\Delta_2 = 2.28$ meV is similar to what has been reported in STM studies. [11,16] However, the first gap has not been detected by STM experiments. This could be explained by a broader distribution of wave vectors for injected electrons in our device configuration, which leads to a probing of the SC gap function $\Delta(\bm{k})$ with the wave vector $\bm{k}$ from different Fermi pockets. We note that such a two-plateau feature with similar energy scale is reproducible in our experiments, as shown in Fig. 4 by two N-S junctions from different devices (See Supplementary Data for more N-S junctions showing the observation of two gaps and a discussion on determining ballistic conduction for obtaining accurate gap energy). The observed smaller plateau could not be explained by the zero-energy surface bound state in *d*-wave superconductors [22] as shown by its different *B*-field dependence (Fig. 2). The two-gap scenario seems to be the most reasonable explanation.

Moreover, we have performed additional diagnostic experiments to verify the superconducting behavior of the microcrystal under study. Fig. 5a presents the differential resistance *R* measured in a standard four-terminal configuration as a function of DC bias current at different temperatures. The data demonstrate a decreasing critical current $I_c$ as temperature increases, and the normal state is reached at $T = 15.4$ K. Comparison between the $I_c$ curves (Fig. 5a) and the AR spectra (Fig. 2b) confirms that the characteristic features in Fig. 2b are indeed related to a superconducting state. Fig. 5b shows $R$ vs. $T$ at zero bias current, revealing a superconducting transition started at $T \sim$ 15K. This is consistent with the temperature dependence of the SC energy gap (Fig. 5c) extracted from the AR data in Fig. 2b. For comparison, we also plot the temperature dependence of the energy gap for conventional BCS superconductors [28] (Fig. 5c). At temperatures lower than 0.5 $T_c$, our data show a weak temperature dependence for the gap, similar to the BCS behavior. In theory, [29] for two coupled Fermi pockets, the resultant two gaps should vanish at the same $T_c$, although a long tail may present in the $T$-dependence of the smaller gap. In our data, the smaller gap is not obvious above $T \sim 9.8$ K, but its existence (with a small or even negligible gap value) before the reach of $T_c = 15$ K cannot be ruled out. We discuss below the possibility of two SC gaps based on the electronic structure of $Fe_{1+y}Te_{1-x}Se_x$. According to a density functional calculation [10], $Fe_{1+y}Te_{1-x}Se_x$ should have at least two hole Fermi pockets at the $\Gamma$ point and two electron Fermi pockets at the M point of the Brillouin zone, since FeSe has two hole (electron) Fermi pockets at the $\Gamma$ (M) point while FeTe has an additional hole pocket at the $\Gamma$ point. It has been suggested that interband coupling between electron and hole pockets is responsible for the superconductivity [13]. One can then address the possible gap structure qualitatively based the theory of Suhl *et al.* [29] by considering just one pair of coupled electron and hole pockets (assuming that the other similar pair of coupled electron and hole pockets lead to superconductivity with similar energy scale). As specifically pointed out in Ref. [29], even with pure interband coupling between two Fermi pockets responsible for superconductivity, one should still expect two SC gaps with different energy scale in general. Therefore, the scenario of two SC gaps resulting from multiple Fermi pockets is feasible (although one cannot rule out the possibility of more SC gaps which might not be clearly resolvable in the experimental AR spectra).

In summary, with uniquely designed multi-terminal N-S devices, we have demonstrated a general experimental approach to point-contact spectroscopy, and employed it to unveil the existence of two SC gaps in $Fe_{1+y}Te_{1-x}Se_x$ and their dependence on temperature and magnetic field. This approach opens up new opportunities to study gap structures in superconductors and elemental excitations in solids.

**Acknowledgements:** We thank Prof. Pei-Herng Hor for helpful discussion and access to experimental facilities.

References

[1] Kamihara Y, Watanabe T, Hirano M and Hosono H 2008 *Journal of the American Chemical Society* **130** 3296.
[2] Ren Z A *et al.* 2008 *Chinese Physics Letters* **25** 2215.
[3] Chen X H, Wu T, Wu G, Liu R H, Chen H and Fang D F 2008 *Nature* **453** 761.
[4] Chen G F, Li Z, Wu D, Li G, Hu W Z, Dong J, Zheng P, Luo J L and Wang N L 2008 *Physical Review Letters* **100** 247002.
[5] Wang C *et al.* 2008 *Epl* **83** 67006.
[6] Hsu F C *et al.* 2008 *Proceedings of the National Academy of Sciences of the United States of America* **105** 14262.
[7] de la Cruz C *et al.* 2008 *Nature* **453** 899.
[8] Lumsden M D and Christianson A D 2010 *Journal of Physics-Condensed Matter* **22** 203203.
[9] Yeh K W *et al.* 2008 *Epl* **84** 37002.
[10] Subedi A, Zhang L J, Singh D J and Du M H 2008 *Physical Review B* **78** 134514.
[11] Hanaguri T, Niitaka S, Kuroki K and Takagi H 2010 *Science* **328** 474.
[12] Song C L *et al.* 2011 *Science* **332** 1410.
[13] Mazin, II, Singh D J, Johannes M D and Du M H 2008 *Physical Review Letters* **101** 057003.
[14] Hu J, Liu T J, Qian B, Rotaru A, Spinu L and Mao Z Q 2011 *Physical Review B* **83** 134521.
[15] Park W K, Hunt C R, Arham H Z, Xu Z J, Wen J S, Lin Z W, Li Q, Gu G D and Greene L H 2010 *arXiv:1005.0190v1*
[16] Kato T, Mizuguchi Y, Nakamura H, Machida T, Sakata H and Takano Y 2009 *Physical Review B* **80** 180507.
[17] Blonder G E, Tinkham M and Klapwijk T M 1982 *Physical Review B* **25** 4515.
[18] Naidyuk Y G and Yanson I K *arXiv:physics/0312016v1*
[19] Duif A M, Jansen A G M and Wyder P 1989 *Journal of Physics-Condensed Matter* **1** 3157.
[20] Daghero D and Gonnelli R S 2010 *Superconductor Science & Technology* **23** 043001.
[21] Chen T Y, Tesanovic Z, Liu R H, Chen X H and Chien C L 2008 *Nature* **453** 1224.
[22] Deutscher G 2005 *Reviews of Modern Physics* **77** 109.
[23] See Supplementary Data for experimental methods, correlation between the critical current and the differential conductance dip, and a discussion on the determination of ballistic conduction in N-S junctions.
[24] Strijkers G J, Ji Y, Yang F Y, Chien C L and Byers J M 2001 *Physical Review B* **63** 104510.
[25] Sheet G, Mukhopadhyay S and Raychaudhuri P 2004 *Physical Review B* **69** 134507.
[26] Plecenik A, Grajcar M, Benacka S, Seidel P and Pfuch A 1994 *Physical Review B* **49** 10016.

[27] Dynes R C, Narayanamurti V and Garno J P 1978 *Physical Review Letters* **41** 1509.
[28] Giaever I and Megerle K 1961 *Physical Review* **122** 1101.
[29] Suhl H, Matthias B T and Walker L R 1959 *Physical Review Letters* **3** 552.

Figure Captions

Figure 1. (a) Schematic diagram of a superconductor microcrystal lying on top of three normal metal electrodes and a circuit designed for obtaining Andreev reflection spectra for the target N-S junction between the $I_+$ ($V_+$) electrode and the superconductor. (b) Optical microscope image of a device with a suspended microcrystal of $Fe_{1+y}Te_{1-x}Se_x$ bridging six parallel metal electrodes (35 nm Pd/ 5 nm Cr) spaced ~ 500 nm apart. The four center electrodes (labeled as 2-5) are designed to be 1 µm wide while the two outside electrodes (labeled as 1 and 6) are 4 µm wide.

Figure 2. (a) Normalized differential conductance $dI/dV$ (symbols) versus bias voltage $V$ at temperature $T = 240$ mK and magnetic field $B = 0$ for the N-S junction between electrode 4 and the superconductor of Fig. 1b, and a fitting (solid line) to the data by the BTK theory (see main text). Raw data of $dI/dV$ vs. $V$ measured (b) at different temperatures with $B = 0$ and (c) under different magnetic fields perpendicular to the substrate at $T = 240$ mK for the same N-S junction.

Figure 3. (a) Measured $dI/dV$ across the N-S junction 4 (right axis) and standard four-terminal differential resistance $R$ (left axis) for the superconductor crystal of Fig. 1b as a function of DC bias current $I$ at temperature $T = 240$ mK under magnetic field $B = 0$, showing a concurrent occurrence of the $dI/dV$ dip in the AR spectrum and the critical current $I_c$ (characterized by a peak in the differential resistance preceding the normal state). Measurement configurations are labeled by the electrode numbers in a sequence

of $I_+$ - $V_+$ - $V_-$ - $I_-$. Same $I_+$ and $I_-$ terminals were used to keep the electric field distribution inside the superconductor crystal identical in both measurements. (b) Schematic current-voltage (I-V) curve (top) for a N-S junction showing a linear I-V curve at low bias with higher conductance (due to Anreev reflection) and a linear I-V curve at high bias with lower conductance (after the normal state is reached), both extending through the origin following Ohm's law; and the corresponding differential conductance vs. voltage (bottom) where a dip occurs in the transition region. (c) $dI/dV$ vs. $V$ for the N-S junction of (a) measured with different combinations of the $V_-$ and $I_-$ terminals at $T = 240$ mK and $B = 0$.

Figure 4. Andreev reflection spectra of (a) the N-S junction of Fig. 2 and (b) another junction showing similar energy scale of two gaps.

Figure 5. (a) Standard four-terminal differential resistance $R$ as a function of DC bias current $I$ at different temperatures; and (b) $R$ as a function of temperature $T$ at zero bias current for the microcrystal of Fig. 1b. (c) Energy gaps as a function of temperature extracted from experimental data (squares) and the BCS values adopted from Ref. 28 (circles). The energy gap values are obtained from the experimental Andreev reflection data by taking the full width at half maximum of the corresponding $dI/dV$ plateaus as $2\Delta$. Full error bars of gaps are determined from the range of bias $V$ where $dI/dV$ changes from 30% to 70% of the relevant plateau height. The BCS values are used for a gap $\Delta_2 = 2.28$ meV corresponding to a $T_c$ of 15.0 K.

Figures

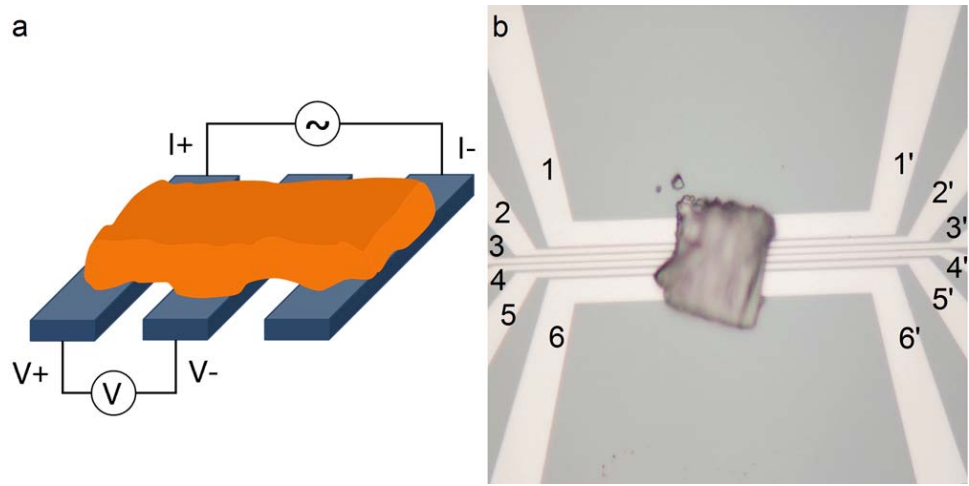

Fig. 1

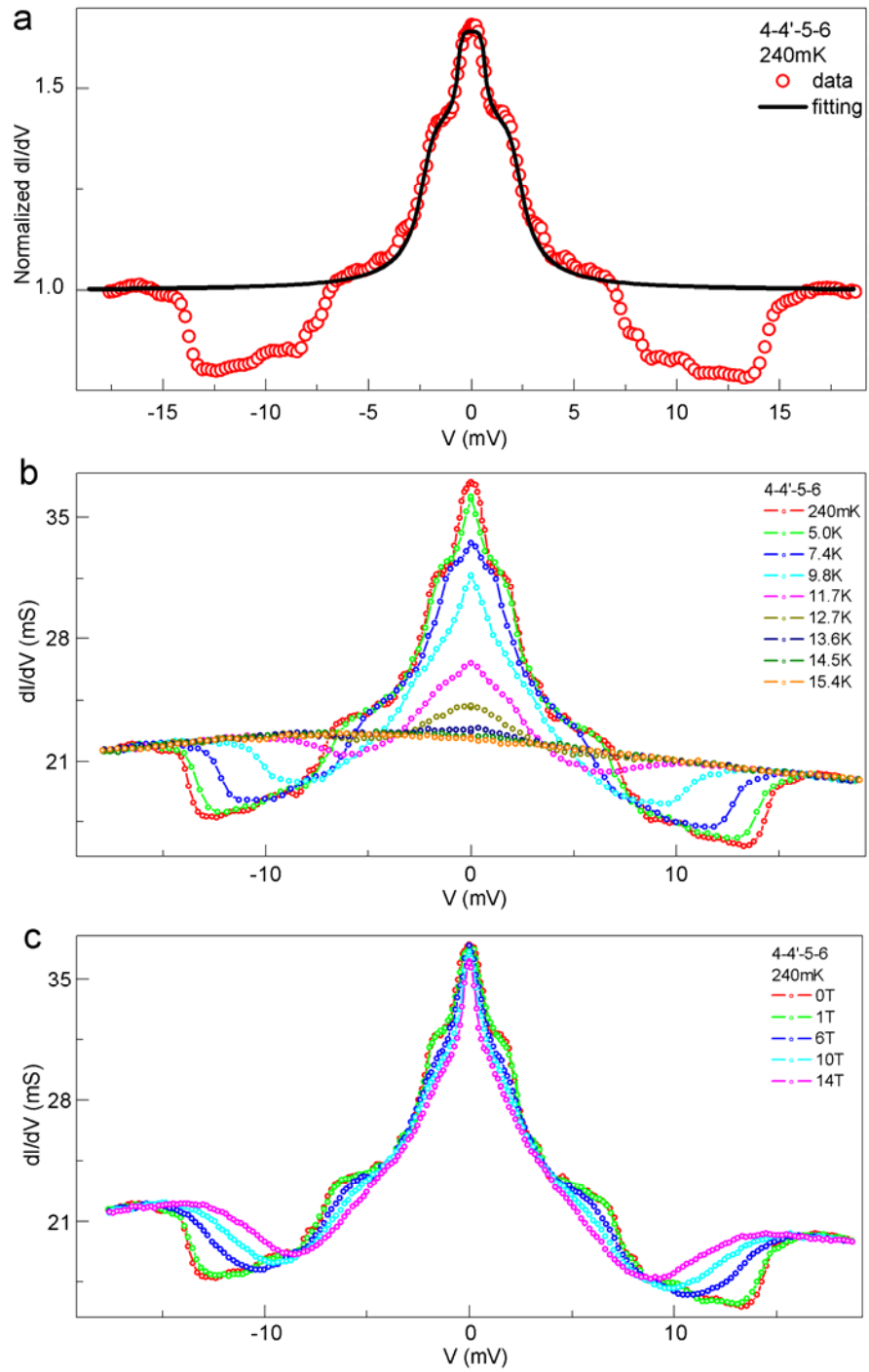

Fig. 2

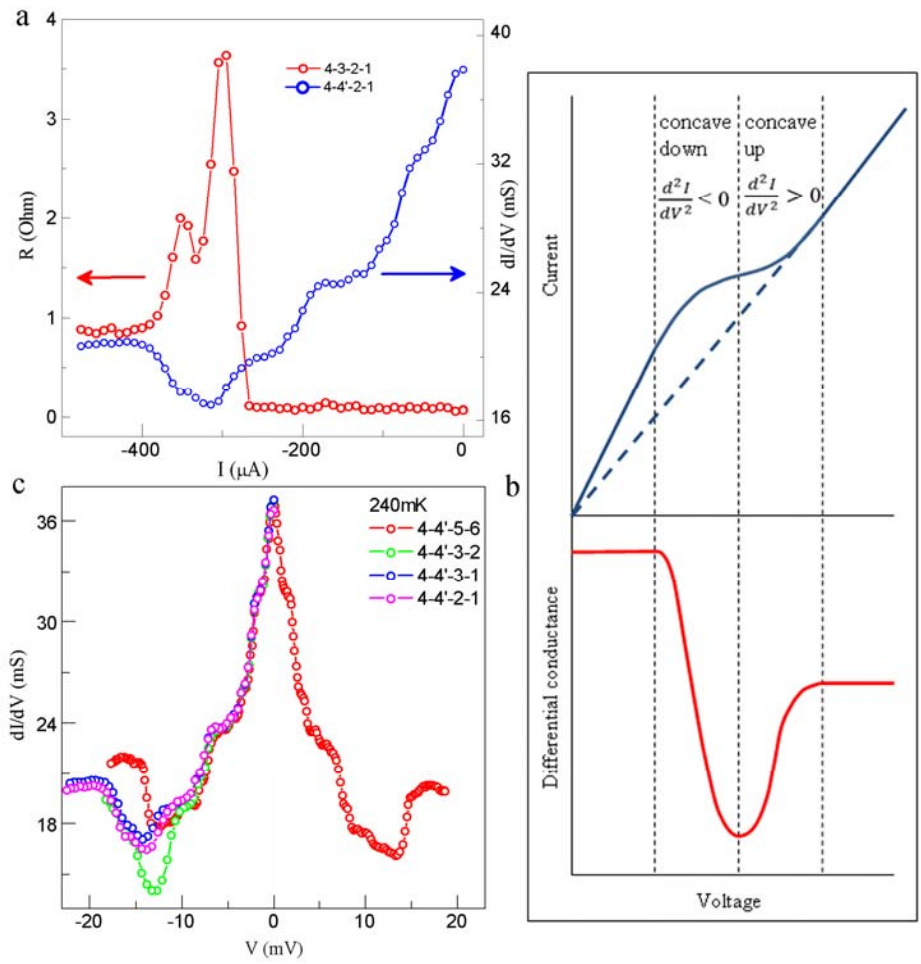

Fig. 3

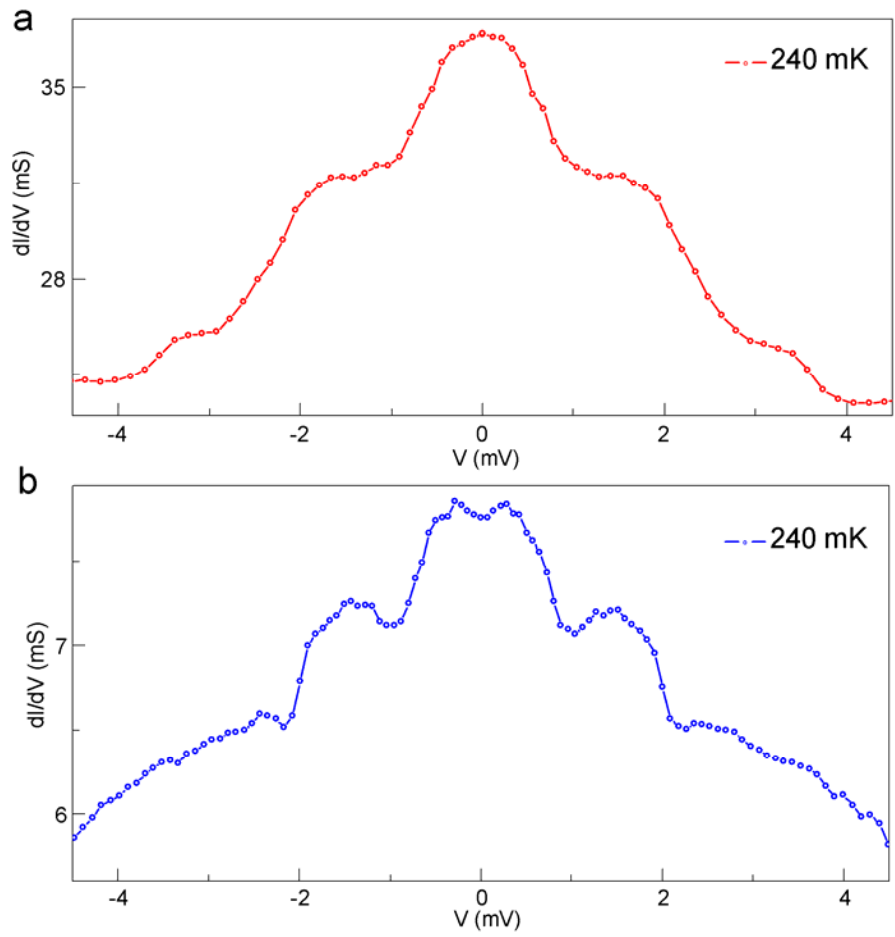

Fig. 4

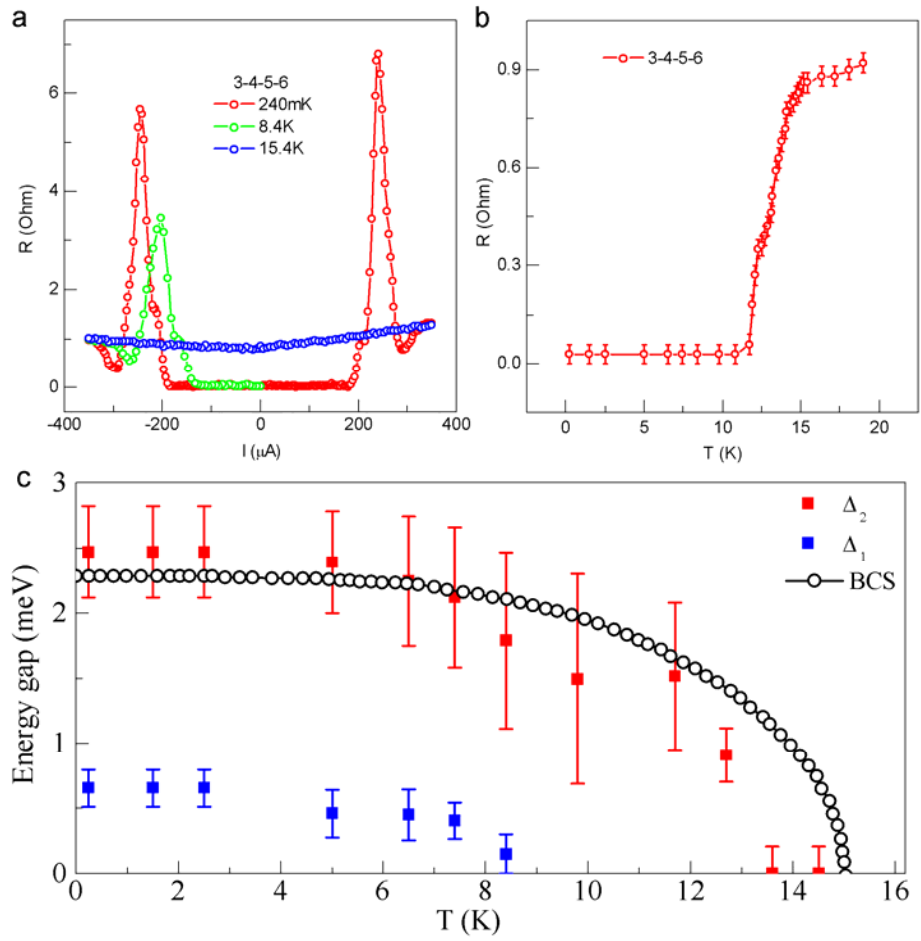

Fig. 5

Supplementary Data for:

# Observation of multiple superconducting gaps in $Fe_{1+y}Te_{1-x}Se_x$ via a nano-scale approach to point-contact spectroscopy

Haibing Peng*, Debtanu De, Zheng Wu, Carlos Diaz-Pinto

Department of Physics and the Texas Center for Superconductivity, University of Houston, 4800 Calhoun Rd, Houston, Texas 77204-5005

* Corresponding author: haibingpeng@uh.edu

## I. Experimental Methods

We start with single-crystal bulk materials of $FeTe_{0.5}Se_{0.5}$ (nominal composition targeted for crystal growth) showing an onset $T_c$ of 14.2 K, as determined by measurements of four-terminal resistivity and AC magnetic susceptibility. The actual composition of the materials is close to $Fe_{1.09}Te_{0.43}Se_{0.57}$ as determined by energy dispersive X-ray spectroscopy. Micro-meter scale crystals are formed by mechanical cleaving of the bulk crystal either with a surgery knife or a pair of mortar and pestle. Such micro-crystals are then placed on $SiO_2$/Si substrates with lithographically patterned electrodes and immediately transferred into a vacuum chamber. We usually select those shinning micro-crystals in the order of 10 μm in lateral dimensions under an optical microscope. Inside the vacuum chamber, a probe tip 25 μm in radius attached to a micro-manipulator is used to place a target superconductor micro-crystal onto multi-terminal normal metal electrodes pre-patterned on the substrate. Typically, we have six parallel electrodes (35 nm Pd/ 5 nm Cr) spaced ~500 nm apart which are patterned by electron-beam lithography on top of a 200 nm thick $SiO_2$ layer thermally grown on Si wafers. Due to the uneven surface of superconductor micro-crystals, the actual N-S contact should be treated as a point contact with a size much less than the designed electrode width (~1μm). Nevertheless, the width of the electrodes does offer certain control of the actual point contact size. To improve the electrical contacts in some N-S junctions, we apply a short voltage pulse (up to ~20 volts) to anneal the device in vacuum. The devices are then loaded into a $He^3$ fridge for low temperature measurement. Magnetic fields are applied perpendicular to the substrate with a superconducting magnet inside a cryostat. The critical temperature and the critical current of the SC microcrystal are characterized in a standard four-terminal configuration. To obtain the AR spectra for a particular N-S junction, we employ a three-terminal circuit (Fig. 1a of the main text), where a small AC current at a frequency of 503 Hz is superimposed to a DC bias current and applied between the $I_+$ and $I_-$ electrodes. Both the DC voltage drop and AC voltage drop across the N-S junction are monitored between the $V_+$ and $V_-$ electrodes. The $I_+$ and $V_+$ terminals are connected to the opposite ends of the same electrode so that the measured voltage drop is exactly from the target N-S junction by excluding the electrode resistance in series.

## II. Correlation between the critical current $I_c$ and the differential conductance ($dI/dV$) dip in Andreev reflection spectra.

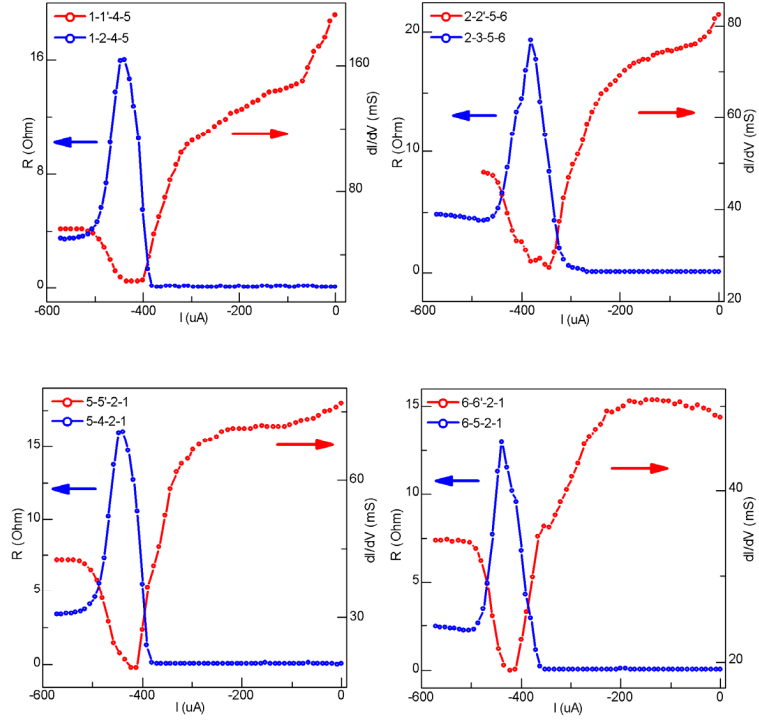

**Fig. S1** Measured $dI/dV$ across different N-S junctions for the device of Fig. 1b in the main text (right axis) and relevant standard 4-terminal differential resistance $R$ for the corresponding sections of the superconductor crystal (left axis) as a function of the DC bias current at temperature $T = 240$ mK under zero magnetic field. Same $I_+$ and $I_-$ terminals were used to keep the electric field distribution inside the superconductor crystal identical for both measurements so that the critical current occurs at the same value. Measurement configurations are labeled by the electrode numbers in a sequence of $I_+$ - $V_+$ - $V_-$ - $I_-$.

## III. Empirical rules for determination of ballistic conduction in N-S junctions

Here we comment on the issue of determining the conduction regime in N-S junctions, important for providing spectroscopic, energy-resolved information on the SC gap. In principle, the gap energy can only be measured accurately in either the ballistic regime with the actual point contact radius $a$ much less than the mean free path $l$, or in the diffusive regime with no inelastic scattering (despite introducing a non-ideal effect of reducing the AR ratio). In traditional AR method [see, Ref. 4 of the main text], the Sharvin formula is used to derive a commonly adopted (but crude) criteria for ballisticity: $R_N >> 4\rho/3\pi l$, with $R_N$ the normal state resistance of the point contact and $\rho$ the bulk material resistivity. This leads to an empirical rule of selecting $R_N$ in a range of 10 – 100 Ω for most materials. However, the measurement of $R_N$ is complicated due to a spreading resistance of the electrodes in series with the N-S point contact in the needle-anvil method. Moreover, the existence of a surface layer renders the use the bulk values of $\rho$ and $l$ inaccurate for estimating a satisfactory $R_N$ range, causing problems in obtaining accurate and reproducible energy gap values. Here, relying on the multi-terminal device design, we propose an extra empirical rule as the criteria of ballisticity in a N-S junction: the measured SC gap features should remain the same regardless of the selection of $I_-$ terminals. An example can be seen in Fig. 3c of the main text, where the two center plateaus do not change as the $V_-$ and $I_-$ terminals are varied, suggesting a ballistic conduction in the N-S junction. This empirical rule is reasonable since the $I_+$ and $I_-$ terminals determine the electric field distribution across the device and inelastic scattering causes energy loss (or voltage drop) inside the N-S junction. Therefore, for N-S junctions in the thermal regime with inelastic scattering, the measured voltage drop does not reflect the incoming electron energy in the AR process and the measured gap plateau features can vary with different $I_-$ terminals. Fig. S2 shows such an example N-S junction in thermal regime.

We also note that in experiments, for N-S junctions with higher normal-state resistance such as those shown in Fig. 4 of the main text, the energy of the observed two gaps is reproducible, indicating ballistic conduction and an accurate determination of the gap energy. For some N-S junctions with lower normal-state resistance (Fig. S3), the two-gap feature is still observable, but the occurrence of the critical-current induced dip may appear near the edge of the larger gap, indicating a point-contact size larger than the coherence length of the superconductor and thus an inaccurate determination of the gap energy [this effect has been discussed in Ref. 4 of the main text].

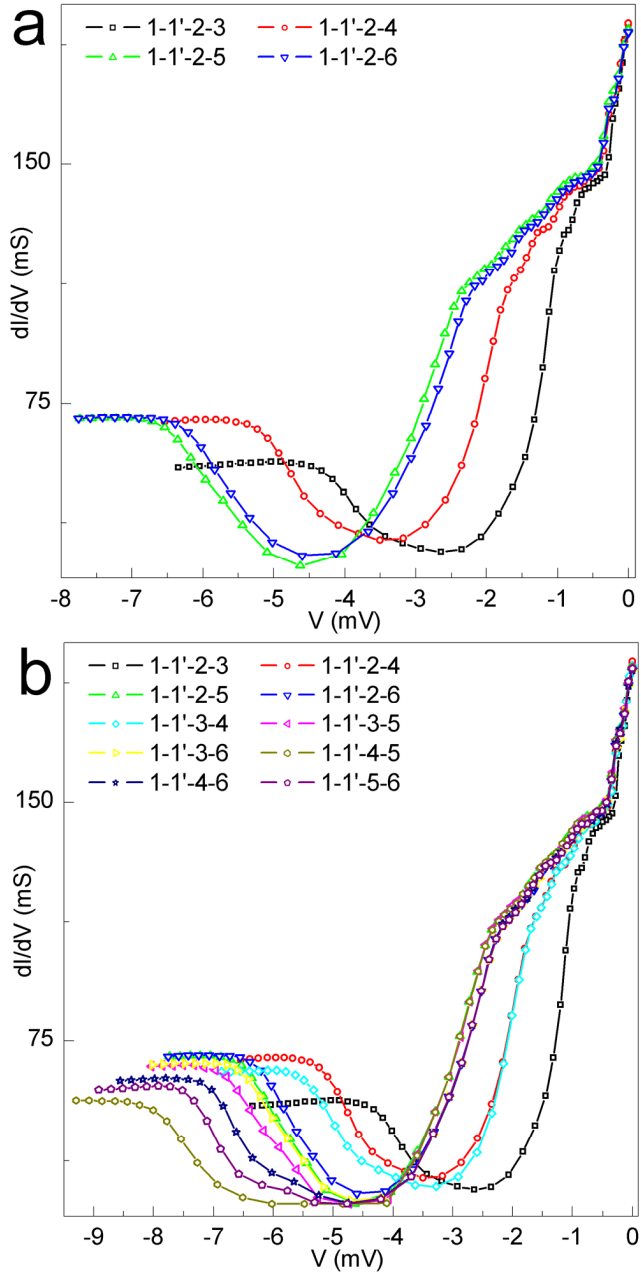

**Fig. S2** (a) *dI/dV* vs. *V* for the junction between electrode 1 and the superconductor of Fig. 1b in the main text, measured with different $I_-$ terminals at temperature $T = 240$ mK and $B = 0$. Electrode configurations are labeled in a sequence of $I_+$ - $V_+$ - $V_-$ - $I_-$. The plateaus vary as the $I_-$ electrode is changed, indicating a thermal regime with inelastic scattering in the point contact. In addition, the selection of the $I_+$ and $I_-$ terminals affects the critical current and thus changes the position of the corresponding *dI/dV* dip preceding the normal state. In some measurements, the $I_c$ is reached even at $V < \Delta/e$. (b) *dI/dV* vs. *V* for the same junction measured with all possible combinations of $V_-$ and $I_-$ terminals. The selection of $V_-$ terminals affects the nominal normal state *dI/dV* values, since the measured voltage drop between the $V_+$ and $V_-$ electrodes includes a contribution from the superconductor crystal once the normal state is reached. Note that the measured zero bias *dI/dV* value is more than twice of the nominal normal state value, because with a fixed AC driving current *dI* the measured AC voltage *dV* includes a voltage drop on the crystal when the normal state is reached and thus the measured normal state *dI/dV* is smaller than the actual one across the N-S junction.

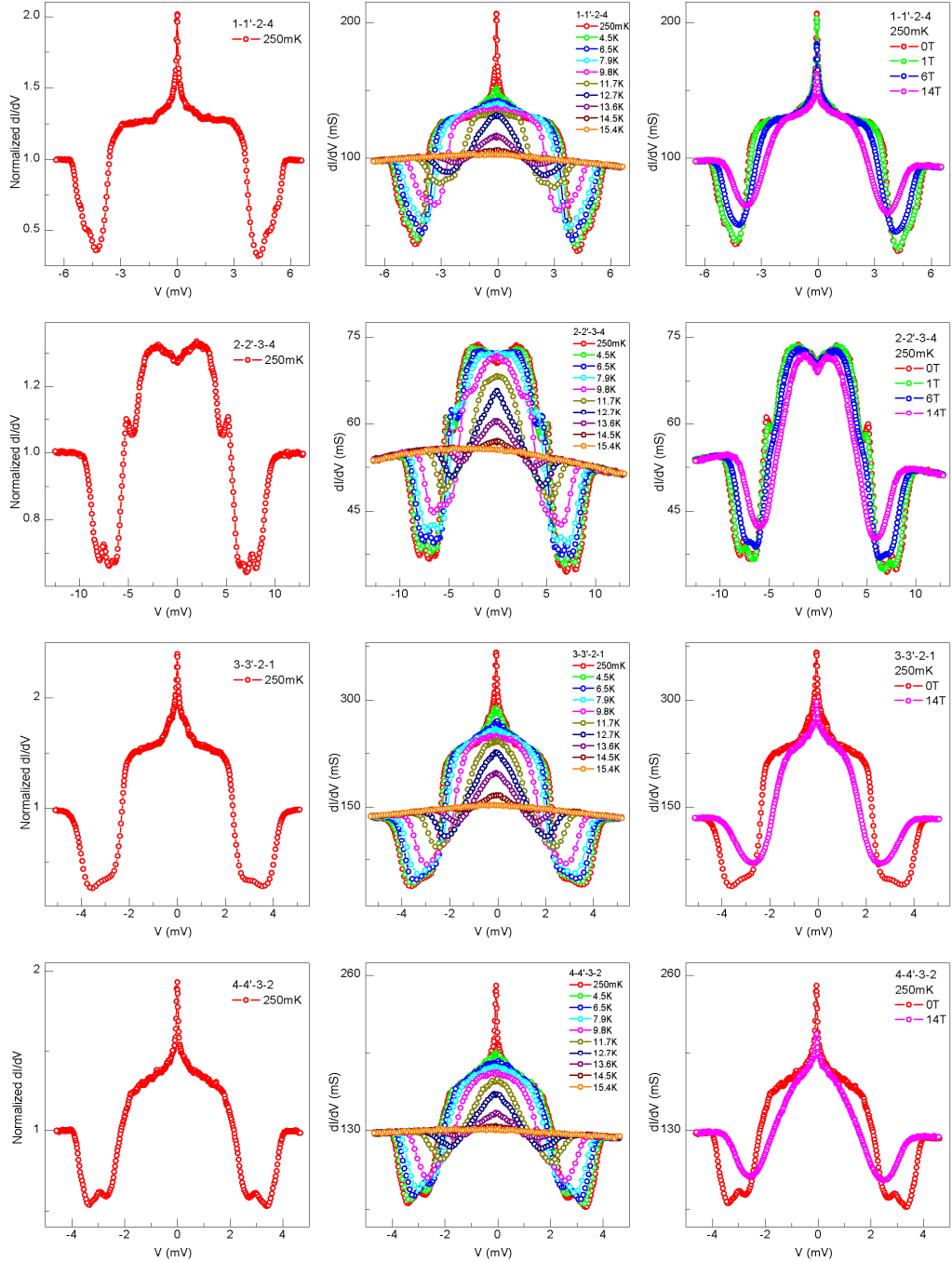

**Fig. S3** The AR spectra for four N-S junctions in another device measured at various temperatures and magnetic fields. Shown in each row are the normalized $dI/dV$ at $T$ = 250 mK and $B$ = 0 (left), raw data of $dI/dV$ vs. $V$ at different temperatures (center), raw data of $dI/dV$ vs. $V$ under magnetic fields (right). For the N-S junction of the second row, a two-gap scenario is still valid since the $dI/dV$ dip at zero bias can be attributed to a higher barrier strength $Z$ for the smaller energy gap which is not observable above $T$ = 9.8 K.

5